# Take Care: A Study on Usability Evaluation Methods for Children

Mohammadi Akheela Khanum
Faculty of Computer Science
PAHER University
Rajasthan, India
e-mail: akheela.khanum@gmail.com

Munesh Chandra Trivedi
DIT School of Engineering
Greater Noida, India
e-mail:munesh.trivedi@gmail.com

*Abstract*: Nowadays whenever a user buys any gadget, apart from the price his focus would also be on how easy is the functionality of the gadget. This means users are more focussed towards the usability of the gadget. Usability of a system is defined as the easiness with which the user can use the system to perform the required tasks. Therefore, during the system development cycle, usability evaluation is performed. Usability evaluation involves testing a specific system by involving a population of the target users. The results of usability evaluations can be incorporated into the system design in order to make the system usable and likeable by the target users. Therefore, this study set to explore usability evaluations methods for children in order to analyze their roles in the development of technology. Usability evaluation methods which are successfully tested on the adults are investigated to find out how successfully they can also be applied to children. The results of the review indicate that usability evaluation with children is more challenging than with adults. Children display varying behaviours in varying environments. Therefore, usability evaluation with children is more about understanding the children's psychological and behavioural aspects. Strong empirical base is needed to understand the children's behaviour in different contexts and accordingly to choose the appropriate usability evaluation method. The study found that children's logical thinking abilities are not fully developed, so depending on one type of usability evaluation method would not be an appropriate decision.

## I. INTRODUCTION

Human Computer Interaction (HCI) is a well established field since more than 25 years. It deals with the synergy between the human and the technological aspect of interaction. A subfield of HCI called the Child Computer Interaction (CCI) is an emerging field, which deals specifically with interaction design for children. Markopoulos et al. [1] states "Relating to sociology, education and educational technology, connected to art and design, and with links to storytelling and literature, as well as psychology and computing this new field borrows methods of inquiry from many different disciplines". The trend of children using information and communication technologies in their day to day lives is increasing drastically. This increase in the use of interactive technology by the children has urged the technology manufacturers to turn their attention towards this rapidly growing market segment. Children are not just young age individuals; they represent a set of individuals who have their own perception, style, preferences, likes and dislikes. When designing technology for children their preferences should be taken into account. To do so, usability evaluations are performed with the children as the testers of technology. During the early design phases of children technology, usability engineers performs usability testing to uncover usability problems that might creep into the product when set to be used in the real context. The best way to achieve this is to involve the children in the test. Substantial amount of literature has been found wherein children are involved in testing the technology which is designed for them. Incorporating children in usability evaluations is very challenging. It involves ethical concerns as well as concerns relating to the recruitment of the children for the test. The ethical concerns are related to the safety of the children, seeking permission from their parents and ensuring the parents about their children's care during the test. Recruiting the children involves selecting the most appropriate children for the product being tested. This can be done by involving teachers and parents. When testing with children, appropriate usability evaluation methods (UEMs) have to be selected.

### A. Usability Evaluation Methods with Children

Many studies in the past have focused on comparing the applicability of the different UEMs to children which were successfully applied and tested on adults. In this section we focus on the qualitative usability evaluation methods applied to usability testing with children. Qualitative usability evaluation methods produce subjective description of the test results. They do not produce numeric data. These methods can be broadly classified into three categories (1) Introspection (2) Direct Observation and (3) Interviews and Questionnaires. The classification is depicted in Table I below.

Table I. Qualitative Evaluation Methods

| |
|---|
| **Introspection** |
| **Direct Observation**<br>• Simple Observation<br>• Think Aloud<br>• Constructive Interaction |
| **Interviews and Questionnaires** |

*Introspection* is the most common evaluation method. Designer tests the system (or prototype) for potential flaws.
This method is not reliable as it is completely subjective and real users are not involved in the test.
In *direct observation* method the evaluator records the user interactions with the system. This can be done either in a controlled environment-the lab or in the real environment-the field. This method is good in identifying the gross deign or the interface problems. Three approaches of direct observation are (i) simple observation: in order to test the given system the user is given a task to perform, and the evaluator just watches the user. The method is simple but it does not give insight into the user's decision process or attitude. (ii) Think Aloud (TA): it is the most widely used testing method in industry. The test users are asked to speak loudly about what they think is happening, what they are trying to do and why they perform an operation. Thinking aloud can give insight into what the user is thinking. The negative side of this method is users may not feel convenient to speak and perform the tasks simultaneously, especially, children. (iii) Constructive Interaction (CI): In CI two testers work together on a given task. The conversation between the two testers during the test is monitored. Nielsen [2] claims that constructive interaction is preferable over think-aloud when conducting usability evaluations with children. Where children face difficulties in following the instructions for a think-aloud test, constructive interaction comes closer to their natural behavior, since the children work in pairs and collaborate in solving the tasks.

*Interviews* are another way of exploring the user experience during usability evaluations. Post-task interviews can be used to probe more deeply on interesting issues. The post-task interview allows observation and verbalization data to be obtained quickly without analyzing tapes. Post-task interviews can offer benefits at the cost of slightly longer evaluation sessions with children.

In this paper we try to seek the answers to the following questions: (i) what are usability evaluation methods that have been used with children? (ii) Why the usability evaluation methods for adults cannot be used for children? In what follows, section 2 describes the related work, in section 3, 4 and 5 we try to find the answers to the above mentioned research questions .

## II. RELATED WORK

In this section we take a sneak peek into the literature on the usability evaluations involving children as testers of technology. Most of the work we surveyed deals with comparing the different usability evaluation methods.

Donker & Markopoulos [3] studies a comparative assessment of three UEMs namely the Concurrent Think Aloud (CTA), interview and questionnaire. Each of these UEMs requires a different level of verbalization for the children that are performing the evaluation. In order to tests these three evaluation methods, 45 children aged 8-14 years were recruited as the test users. The result indicates that children who think aloud during testing uncover more problems than the children who answer specific questions. However, to elicit verbal comments the children have to be prompted, which can be an indication that children find it difficult to think aloud. Prompting may cause children feel obliged to mention problems to please the experimenter. This could lead to non problems being reported. The result also suggests that girls thinking out loud report more usability problems than boys.

Baauw and Markopoulos [4] conducted a study to compare UEMs. The study involved twenty four children in the age group of 9-11 year, in the usability testing of the computer game- BioMania. The usability evaluation was carried out to test two UEMs namely the TA and post task interview. The results indicate that there was no significance difference between the problems reported by the two genders. The post task interview allows observation data and verbalization data to be obtained on fly without analyzing tapes. Thus, post task interviews can offer practical benefit at the cost of slightly longer sessions. The number of usability problems identified through the two methods was not significant.

Markopoulos and Bekker [5] presented a framework for characterizing comparative studies of usability testing methods with respect to their appropriateness for children. They found that the ability to verbalize problems in interactions depends on: the ability of translating experiences into verbal statements, on their knowledge of the language and on prior experiences in speaking up to adults. They found that compound tasks and abstract tasks formulations could pose problems to children, as their abstract and logical thinking abilities are not yet fully developed and they are not skilled in keeping multiple concepts simultaneously in mind. The results also indicate that think aloud helps generate more problems reports than questionnaires and interviews.

Vermeeren et al., [6] conducted a study on the use of post task interviewing evaluation technique with 6-8 years old children. The results show that children overall were fairly good at answering the questions. The negative side effects of applying the technique on the outcome of the usability test are minor. Further, the study suggests applying such technique to uncover extra data about possible causes for interaction difficulties. Also to limit the questions by only asking detailed questions about those parts of the design that needs extra attention.

Al Wabil et al., [7] in their study ten children in the age group of 8-13 years were involved in the evaluation of two different websites, one educational website and one entertainment website. The Retrospective Think Aloud (RTA) protocol was used along with the Eye gaze replay in post session interviews. Findings show that the stimulated RTA protocol with eye tracking is effective in eliciting exploratory information about what children attend to in usability evaluations and how they process information and how children arrived at a target element or solution.

Read and Fine [8] in their study explores four known concerns with using survey methods. These concerns are: (i) statisfycing and optimizing (ii) suggestibility (iii) specific question format, and (iv) language effects. The study suggest that because the researchers and developers of interactive product are generally not specialists in survey design and so invariably produce questions and suggested answers that are far from perfect. The study suggests that the survey methods for children have some inherent difficulties. Such methods should be discouraged when applying to the children.

A study by Als, Jensen and Skov [9] presents an experiment that compares TA and Constructive Interaction (CI). Sixty children in the age group of 13 and 14 years were used as test subjects. Testing was carried out in three different setups. These are (i) the individual testers, (ii) the acquainted dyads, and (iii) the non-acquainted dyads. All the acquainted dyads were familiar with each other and studied in the same class. The non-acquainted dyads attend different schools. The results indicate that CI did not necessarily facilitate natural think aloud as the dyads tended to talk-aloud and not think-aloud. Dyads configuration in CI influenced the children's behavior and assessment of the testing situation according to their acquaintance. Gender issues might play important roles in the configuration of the dyads in CI.

Edwards and Benedyk [10] propose a study that assesses three usability evaluation methods, Active Intervention, Peer Tutoring and Cross-Age Tutoring. Testing was carried out with children aged 6-8 years within a school setting, using an interactive educational multimedia product. Cross-Age Tutoring elicited significantly fewer comments than the other two methods and 'plan' comments were significant rarer than 'action' and 'perception and cognition' comments. In terms of the suitability of these evaluation methods for child participants, and context of use in this particular setting, Peer Tutoring appears to have the most potential.

van Kesteren et al., [11] proposed an exploratory study to look at the children's ability to provide verbal comments in usability evaluation sessions. Six evaluation methods were applied to test an interactive toy by children aged 6 and 7 years old. The results show that most verbal comments were gathered during Active Intervention sessions by asking children questions during tasks. Co-Discovery sessions were less successful, because children did not collaborate very well. Children also provided useful comments in the RTA and Peer Tutoring sessions. They could reflect on their actions at the end of retrospection sessions, and were able to teach other children how to interact with the toy in Peer Tutoring sessions. Another study by AlShumait, AlOsaimi and AlFedaghi [12] was carried out to investigate the effectiveness of five survey techniques for the evaluation of the usability of e-learning programs for 5 and 6 years old children. Smileyometer, Best/Worst Activity table and Again/ Again Table have proven to be more reliable survey techniques used with the children, than the "WordBox" and "Remembering".

### III. CHILDREN AND BEHAVIOUR

Behavior of the test participants is affected by the context in which the usability evaluation takes place [13]. Context has been defined differently by different people. Ryan et al [14] define context as the user's location, environment, identity and time. Hull et al [15] defines context to include the entire environment aspects of the current situation. One of the widely accepted theories of human behavior is credited to Roger G. Barker. His theory of behavior settings can be used a tool to study the human behavior. While the theory has strong empirical base, research on it is limited. Barker and his colleagues continuously collected empirical data from a small town in Kansas with less than 2000 people from 1947 through 1972 based on which he developed the theory of behavior settings. Behavior setting theory proposes that there are specific, identifiable units of the environment, the physical and social elements, which are combined into one unit, and have very powerful influences on human behavior [16]. Behavior setting consists of the behavior aspect and the milieu-the settings, the behavior is circumjacent to the settings. That is, the behavior occurs in the settings and has a strong influence of the settings. Continuous records of the behavior of individual children show that the ever-changing aspect of the child's stream of behavior is one of its most striking features [17]. A close interrelation of settings and people as seen through the Barker's theory of behavior settings could be an indication that context is important and plays a vital role in influencing the results of usability evaluations. Therefore, when testing with children impact of the physical surroundings on the children's behavior should also be accounted.

### IV. USABILITY GUIDELINES FOR TESTING WITH CHILDREN

In order to involve children in the usability evaluation process, it is very essential to make a clear plan on the aims and objectives of the test. The type of product under the test influences the selection procedure of the children. Some of the studies in literature have been found to give very useful guidelines for testing with children. Children older than 14 years of age will likely behave as adults in a testing situation and should be treated accordingly [18].

1. Preschool (ages 2 to 5 years)

Children in this age group have a lower concentration period and would not be able to focus constantly on one object. They may try to impress the adults by showing what they can do on computers without any help. Children in this age group are too young to clearly express their satisfaction levels [18].

2. Elementary School (ages 6 to 10 years)

Usability testing involving children in this age range is easier to include in software usability testing. They are able to follow a task with a higher attention span. They can describe their satisfaction levels properly [18].

3. Middle School (ages 11 to 14 years)

This group is the easiest and mostly used in usability testing. They may be somewhat familiar with the use of the computers. They may be able to "think aloud" during the session, while others may be self-conscious [18].

In what follows we describe some the general guidelines for usability testing with children. These guidelines are categorized in 4 chronological orders of a routine test: Set-up and Planning, Introduction, During the Test and Finish up [18].

*A. Set-up and Planning*

- Make the lab a little more child-friendly but don't overdo.
- Use the input devices that the children are familiar with.
- Keep the laboratory equipment such that they don't distract the children's attention. For example avoid facing children directly toward the video camera or a one-way mirror.
- Avoid using the same sequence of the tasks when planning a series of tasks. This may avoid children from getting bored.
- Select the children who have the required amount of experience to use the computers. This will help focus on the test rather teaching them how to use the computer.
- Try giving them on-time breaks.
- Avoid using the children who are experts in using the computers (unless they are your target audience).
- It is not a good idea to use your own or colleagues children as participants in usability testing.

*B. Introductions*

- Introducing each other will help in establishing relationship with children. This in turn will reduce the stress of the test.
- Parents should sign the agreements because they are the legal guardians.
- Have a script ready which can be used to introduce the test to all participants in the same way.
- Motivate older children by emphasizing the importance of their role.

- Set children's expectations appropriately for what they will be doing during the usability session.
- Children and Parents can be shown the lab, including the behind one-way mirrors. This can give the children a better sense of control and create a trust on researchers.
- Younger children (up to 7- or 8-year-olds) will need to have the tester in the room with them.
- Younger or shyer children may be uncomfortable alone with the tester.
- If siblings accompany children to a test, they can be made to sit in another area away from the test for the duration of the test.

*C. During the Test*

- Preschool-aged children may need a little warm-up with the computer at the beginning of the test. This can be done by asking them to do some small activities with the computers.
- Break down the tasks into smaller segments than for adults, particularly for complex activities.
- Children at times tend to ask for help if they are not sure what to do. Researchers need to redirect children questions by other questions.
- Try to use close ended questions with children. This will reduce the burden of decision making by children.
- When children lose their attention, they should be gently reminded to pay attention to the computer.
- Some children may struggle to read words or numbers. In such cases the researchers should be trickier and try to motivate them by asking them to guess the answer.
- Encourage children by offering generic positive feedback and telling them how hardly they worked without any help.

*D. Finishing Up*

- Behavioral responses such as frowns, sighs, yawns or turning away from the computer are more reliable indicators than their responses which could be sometimes only to please the adults.
- Older children may be able to give reliable ratings about aspects of the software.
- Rewarding children by commenting on how helpful they were. This will reduce the stress made by the test.
- Children and parents often appreciate a choice of a gift certificate.

V. SUMMARY AND DISCUSSION

The summary of the surveyed work is given in Table II.

Table II. Summary of surveyed Work

| Reference | UEM tested | Gender of testers | Age | Number of test participants | Product tested | Results |
|---|---|---|---|---|---|---|
| [3] | CTA, Interview and Questionnaire | Both | 8-14 | 45 | Semi-educational game | Girls thinking aloud resulted in more problems detection than boys |
| [4] | TA, Post task interviews | Both | 9-11 | 24 | Computer game–Bio Mania | No significance difference between genders and methods were found |
| [7] | RTA | Both | 8-13 | 10 | Entertainment and Educational websites | Stimulated RTA is effective in eliciting verbal comments from children |
| [9] | TA, CI | Both | 13-14 | 60 | Inno-100 mobile phone | Pairing of children in CI had impact on how the children collaborated in pairs |
| [10] | Active intervention, Peer tutoring, Cross-age tutoring | Both | 6-8 | Not defined | Interactive educational multimedia product | Peer tutoring appears to have most potential |
| [11] | TA, RTA, Active intervention, Co-discovery, Peer Tutoring | Both | 6-7 | 7 | Jammin Draw | Co-discovery was less successful , RTA , TA and Peer Tutoring sessions gave useful comments |
| [12] | Survey techniques- Simileyometer, Best/worst activity, Again/again table, Word box, Remembering | Both | 5-6 | 17 | E-learning program by ReDSOFT | Simileyometer, Best/worst activity, Again/again table were more reliable for 5-6 year old children. |

The usability evaluations with children are done by assessing the children's performance with the system under test. Usually more than one UEM is applied and the results are compared to reach at more concrete conclusions. The surveyed literature indicates that children with a minimum age of 5 years and a maximum of 14 years are chosen for the test. The total number of test participants varies, some studies choose very lesser (7) participants and some studies choose higher number (60) of participants. The survey of usability evaluation methods on children reveals that various evaluation methods have been tested on children in the past. Every method requires a different level of verbalization from the children. TA and its variants such as CTA, RTA is found to be the prominent choice for testing with children in the average age groups ranging from 8 to 14 years. TA methods elicit natural verbalization behavior from the children. More usability problems are uncovered with TA than with answering specific questions as in the interviews and questionnaires. Some studies also pointed that girls do well with TA than boys. Children who are introvert may find it difficult to express in TA. Stimulating such children to speak out loud during TA sessions may result in wrong problems being reported. This may be due to the fact that when they are prompted, they may try to speak out just to please the experimenter. Study reported in [9] found that CI does not stimulate natural TA because children tend to talk aloud than think aloud. However, acquainted dyads were more satisfied and displayed less workload in CI as compared with non acquainted dyads. This result does not fully supports Nielsen [19] claim which states that CI is better choice than TA when conducting usability evaluations with children. Other UEMs such as the surveys, interviews and questionnaires may have some inherent limitations. Children answer only those questions that are asked from them. Therefore, it is found that such methods can be applied along with other UEMs to uncover more data which may remain hidden due to interaction difficulties in TA and CI sessions. However, such methods can also be helpful in obtaining the verbalization and observation data without analyzing the recorded sessions.

## VI. CONCLUSION

In this study, we explored how different usability evaluation methods can be applied to children in evaluating interactive systems in general. The study reveals that different usability evaluation methods require a different level of verbalization from the children. Usability evaluation methods which involve only verbalization will be too strict for the children. Therefore, it is found that considering flexible approaches in which the child is allowed to express emotion, thoughts, and opinions in activities rather than direct elicitation as is often used with adults. Children's logical thinking and reasoning capabilities are not fully developed; relying only on one type of UEM may not be a wise decision.

## VII. ACKNOWLEDGEMENT